\documentclass[aps,prc,twocolumn,superscriptaddress,nofootinbib,longbibliography,floatfix,10pt]{revtex4-1}

\usepackage{graphicx}
\usepackage{dcolumn}
\usepackage{bm}
\usepackage{morefloats}
\usepackage{multirow}
\usepackage{amssymb}
\usepackage{amsmath}
\usepackage{xcolor}
\usepackage{fix-cm}
\usepackage{mathptmx} % rm & math
\usepackage[T1]{fontenc}
\usepackage[colorlinks,allcolors=blue]{hyperref}
\makeatletter
\def\NAT@def@citea{\def\@citea{\NAT@separator}}
\makeatother

\begin{document}

\title{Quantal Diffusion Description of Multi-Nucleon Transfers in Heavy-Ion Collisions}
\author{S. Ayik}\email{ayik@tntech.edu}
\affiliation{Physics Department, Tennessee Technological University, Cookeville, TN 38505, USA}
\author{B. Yilmaz}
\affiliation{Physics Department, Faculty of Sciences, Ankara University, 06100 Ankara, Turkey}
\author{O. Yilmaz}
\affiliation{Physics Department, Middle East Technical University, 06800 Ankara, Turkey}
\author{A. S. Umar}
\affiliation{Department of Physics and Astronomy, Vanderbilt University, Nashville, TN 37235, USA}

\date{\today}

\begin{abstract}
Employing the stochastic mean-field (SMF) approach, we develop a quantal diffusion description of the
multi-nucleon transfer in heavy-ion collisions at finite impact parameters. The quantal transport
coefficients are determined by the occupied single-particle wave functions of the time-dependent
Hartree-Fock equations. As a result, the primary fragment mass and charge distribution functions are
determined entirely in terms of the mean-field properties. This powerful description does not
involve any adjustable parameter, includes the effects of shell structure and is consistent with the
fluctuation-dissipation theorem of the non-equilibrium statistical mechanics. As a first application
of the approach, we analyze the fragment mass distribution in $^{48}\mathrm{Ca}+{}^{238}\mathrm{U}$ collisions at
the bombarding energy $E_{\text{c.m.}}=193$~MeV and compare the calculations with the experimental
data.
\end{abstract}

%\pacs{24.10.Jv; 21.30.Fe; 21.65.-f; 26.60.-c}

\maketitle

\section{Introduction}
The renewed interest in the study of ion-ion collisions involving heavy systems is driven partially by the
search for new neutron rich nuclei. For this purpose, a number of  experimental investigations of
multi-nucleon transfer processes have been carried out in heavy-ion collision with actinide targets
near barrier energies~\cite{kozulin2014b,kozulin2014}. Collisions of heavy systems at low
energies predominantly lead to dissipative deep-inelastic collisions and quasi-fission reactions. In
dissipative collisions the part of the bombarding energy is converted into internal
excitations and multi-nucleon transfer occurs between the projectile and target nuclei. In
particular, the quasi-fission reactions of heavy-ions provide an important tool for massive mass
transfer~\cite{toke1985,shen1987,hinde1992,hinde1995,hinde1996,itkis2004,knyazheva2007,hinde2008,nishio2008,kozulin2014,rietz2011,itkis2011,lin2012,nishio2012,simenel2012b,durietz2013,williams2013,kozulin2014,wakhle2014,hammerton2015,prasad2015,prasad2016}.
In quasi-fission colliding ions attach together for a long time but separate without going through a
compound nucleus formation. During the long contact time a substantial nucleon exchange takes place between
projectile and target nuclei. A number of models have been developed for the description of quasi-fission
reaction mechanism in terms of multi-nucleon transfer
processs~\cite{adamian2003,zagrebaev2007,aritomo2009,zhao2016}. The time-dependent
Hartree-Fock (TDHF) theory provides a microscopic alternative for describing heavy-ion reaction
mechanism at low bombarding energies~\cite{nakatsukasa2016,simenel2012,negele1982}. In recent years the
TDHF approach has been extensively used to study the quasi-fission
reactions~\cite{golabek2009,kedziora2010,simenel2012b,wakhle2014,oberacker2014,hammerton2015,umar2006c,umar2015a,sekizawa2016,umar2016}.

The TDHF theory provides a good description for the average values of the collective reaction dynamics,
however the approach is not able to describe the fluctuations of the collective dynamics. In
TDHF studies it is possible to calculate the mean values of neutron and proton drifts. It is also
possible to calculate fragment mass and charge distributions by the particle projection approach
\cite{simenel2010,sekizawa2013,sekizawa2014,sekizawa2017,sekizawa2017a}.  But this description works
best for few nucleon transfer, and therefore dispersions of distributions of multi-nucleon
transfers are severely underestimated in dissipative collisions~\cite{dasso1979,simenel2011}. Much
effort has been devoted to improve the standard mean-field approximation by including fluctuation
mechanism into the description. These include the Boltzmann-Langevin transport approach
\cite{abe1996}, the time-dependent random-phase approximation (TDRPA) approach of Balian and Veneroni
\cite{balian1985,williams2018}, the time-dependent generator coordinate method (TDGCM)~\cite{goutte2005}, and
the stochastic mean-field (SMF) approach~\cite{ayik2008}. The applications of time-dependent density
matrix (TDDM) approach on reaction of heavy system have also been recently reported
\cite{tohyama2002a,assie2009,tohyama2016}. Here, we present applications of the SMF approach on
multi-nucleon transfer reactions~\cite{lacroix2014}.

In essence there are two different mechanism for the dynamics of density fluctuations: (i) The
collisional mechanism due to short-range two nucleon correlations, which is incorporated in to the
Boltzmann-Langevin approach. This mechanism is important in nuclear collisions at bombarding
energies per particle around the Fermi energy, but it does not have sizeable effect at low energies.
(ii) At low bombarding energies, near the Coulomb barrier, the long-range
mean-field fluctuations originating from the fluctuations of the initial state becomes the dominant
source for the dynamics of density fluctuations. In the SMF approach, these mean-field fluctuations
are incorporated into the description of the initial state. The standard mean-field dynamics provides a deterministic description,
in which a well specified initial conditions lead to a definite final state. In contrast, in
the SMF approach the initial conditions are specified with a suitable distribution of the relevant
degrees of freedom~\cite{ayik2008}. An ensemble of mean-field events are generated from the
specified fluctuations of the initial state. In a number of studies, it has been demonstrated that
the SMF provides a good approximation for nuclear dynamics including fluctuation mechanism of the
collective motion~\cite{ayik2008,lacroix2014,lacroix2012,lacroix2013,yilmaz2014a,tanimura2017}.
For these low energy collisions the di-nuclear structure is largely maintained. In this
case, it is possible to define macroscopic variables with the help of the window dynamics. The SMF
approach gives rise to a Langevin description for the evolution of macroscopic variables
\cite{gardiner1991,weiss1999} and provides a microscopic basis to calculate transport coefficients
for the macroscopic variables. In the initial applications, this approach has been applied to the
nucleon diffusion mechanism in the semi-classical limit and by ignoring the memory effects
\cite{washiyama2009b,yilmaz2014,ayik2015a}. In a recent work, from the SMF approach, we were able to
deduce the quantal diffusion coefficients for nucleon exchange in the central collisions of
heavy-ions. The quantal transport coefficients include the effect of shell structure, take into
account the full geometry of the collision process, and incorporate the effect of Pauli blocking
exactly. Recently, we applied the quantal diffusion approach and carried out calculations for the
variance of neutron and proton distributions of the outgoing fragments in the central collisions of
several symmetric heavy-ion systems at bombarding energies slightly below the fusion barriers
\cite{ayik2016}. In another work, we carried out quantal nucleon diffusion calculations and determined
the primary fragment mass and charge distributions for the central collisions of $^{238}\mathrm{U}+{}^{238}\mathrm{U}$
system~\cite{ayik2017}.

In this work, we extend the diffusion description for collisions to incorporate finite impact parameters, and
deduce quantal transport coefficients for the proton and neutron diffusions in heavy-ion collisions.
Since the transport coefficients do not involve any fitting parameter, the description may provide a
useful guidance for the experimental investigations of heavy neutron rich isotopes in the reaction
mechanism. As a first application of the formalism, we carry out quantal nucleon diffusion
calculations for $^{48}\mathrm{Ca}+{}^{238}\mathrm{U}$ collisions at the bombarding energy $E_{\text{c.m.}}=193$~MeV
and determine the primary fragment mass distribution~\cite{kozulin2014b}. In section~\ref{formalism},
we present a
brief description of the quantal nucleon diffusion mechanism based on the SMF approach.
In section~\ref{transport},
we present derivation of the quantal neutron and proton diffusion coefficients. The result of
calculations for $^{48}\mathrm{Ca}+{}^{238}\mathrm{U}$ collisions is reported in section~\ref{results}, and conclusions are
given in section~\ref{conclusions}.

\section{Quantal nucleon diffusion mechanism} 
\label{formalism}
We consider collisions of heavy-ions in which the di-nuclear structure is maintained, such as
in the deep-inelastic collision or quasi-fission reactions. In
this case, when the ions start to touch a window is formed between the colliding ions. We represent the reaction
plane in a collision by $(x,y)$-plane, where $x$-axis is taken in the beam direction in the c.m.
frame of colliding ions. Window plane is perpendicular to the symmetry axis and its
orientation is specified by
\begin{align} \label{eq1}
y-y_{0}(t)=-\left[x-x_{0}(t)\right]\cot\theta(t)
\end{align}
In this expression, $x_{0}(t)$ and $y_{0}(t)$ denote the coordinates of the window center relative
to the origin of the c.m. frame, $\theta(t)$ is the smaller angle between the orientation of the
symmetry axis and the beam direction. For each impact parameter $b$, as described in Appendix A, by
employing the TDHF solutions, it is possible to determine time evolution of the rotation angle
$\theta(t)$ of the symmetry axis. The coordinates $x_{0}(t)$ and $y_{0}(t)$ of the center point of
the window are located at the center of the minimum density slice on the neck between the colliding
ions. In the following, all quantities are calculated for a given impact parameter $b$ or the
initial orbital angular momentum $l$, but for the purpose of clarity of certain expressions, we do
not attach the impact parameter or the angular momentum label to the quantities.

In the SMF approach, the collision dynamics is analyzed in terms of an ensemble of mean-field
events. In each event, we choose the neutron $N_{1}^{\lambda}(t)$ and proton $Z_{1}^{\lambda}(t)$
numbers of the projectile-like fragments as independent variables, where $\lambda$ denotes the
event label. The neutron and proton numbers can be determined at each instant by integrating the
neutron and proton densities over the projectile-like side of the window for each event $\lambda$ by
employing the expression,
\begin{align} \label{eq2}
\left(\begin{array}{c} {N_{1}^{\lambda}(t)} \\ {Z_{1}^{\lambda}(t)} \end{array}\right)=\int & d^{3} r\;\Theta\left[(x-x_{0})\cos\theta+(y-y_{0})\sin\theta\right]\nonumber\\
        &\times\left(\begin{array}{c}{\rho _{n}^{\lambda}(\vec{r},t)} \\ {\rho_{p}^{\lambda}(\vec{r},t)} \end{array}\right).
\end{align}
Here, the quantity
\begin{align} \label{eq3}
\rho _{\alpha}^{\lambda}(\vec{r},t)=\sum_{ij\in\alpha}\Phi_{j}^{*\alpha }(\vec{r},t;\lambda)\rho_{ji}^{\lambda}\Phi_{i}^{\alpha}(\vec{r},t;\lambda)
\end{align}
denotes the neutron and proton number densities for the event $\lambda$ of the ensemble of the
single-particle density matrices. Here and in the rest of the article, we use the notation
$\alpha=n,p$ for the proton and neutron labels. According to the main postulate of the SMF approach,
the elements of the initial density matrix have uncorrelated Gaussian distributions with the mean
values $\overline{\rho_{ji}^{\lambda}}=\delta_{ji} n_{j}$ and the second moments determined by,
\begin{align} \label{eq4}
\overline{\delta\rho_{ji}^{\lambda}\delta\rho_{i'j'}^{\lambda}}=\frac{1}{2}\delta_{ii'}\delta_{jj'}\left[n_{i}(1-n_{j} )+n_{j}(1-n_{i})\right]
\end{align}
where $n_{j}$ are the average occupation numbers of the single-particle wave functions at the
initial state. At zero initial temperature, the occupation numbers are zero or one, at finite
initial temperatures the occupation numbers are given by the Fermi-Dirac functions. Here and below,
the bar over the quantity indicates the average over the generated ensemble. In each event the
complete set of single-particle wave functions is determined by the TDHF equations with the
self-consistent Hamiltonian $h(\rho ^{\lambda})$ of that event,
\begin{align} \label{eq5}
i\hbar\frac{\partial}{\partial t}\Phi_{i}^{\alpha}(\vec{r},t;\lambda)=h(\rho^{\lambda})\Phi_{i}^{\alpha}(\vec{r},t;\lambda).
\end{align}
The rate of changes the neutron and the proton numbers of the projectile-like fragment are given by,
\begin{align} \label{eq6}
\frac{d}{dt}\left(\begin{array}{c} {N_{1}^{\lambda } (t)} \\ {Z_{1}^{\lambda } (t)} \end{array}\right)=&\int d^{3} r\;\delta (x')\dot{x}' \left(\begin{array}{c} {\rho _{n}^{\lambda } (\vec{r},t)} \\ {\rho _{p}^{\lambda } (\vec{r},t)} \end{array}\right)\nonumber\\
&+\int d^{3} r\;\Theta (x')\frac{\partial }{\partial t} \left(\begin{array}{c} {\rho _{n}^{\lambda } (\vec{r},t)} \\ {\rho _{p}^{\lambda } (\vec{r},t)} \end{array}\right),
\end{align}
where $x'=\hat{e}\cdot (\vec{r}-\vec{r}_{0})$,
$\dot{x}'=\vec{\theta}\cdot[\hat{e}\times(\vec{r}-\vec{r}_{0})]-\hat{e}\cdot\dot{\vec{r}}_{0}$ with
$\vec{r}_{0}$ and $\dot{\vec{r}}_{0}$ as the position and velocity vectors of the center of the
window plane in the c.m. frame. Here and below, $\hat{e}$ denotes the unit vector along the symmetry
axis with components $\hat{e}_{x}=\cos \theta$ and $\hat{e}_{y}=\sin\theta$. Using the continuity
equation,
\begin{align} \label{eq7}
\frac{\partial }{\partial t}\rho_{\alpha}^{\lambda}(\vec{r},t)=-\vec{\nabla}\cdot\vec{j}_{\alpha }^{\lambda}(\vec{r},t),
\end{align}
we can express Eq.~\eqref{eq6} as,
\begin{align} \label{eq8}
\frac{d}{dt} \left(\begin{array}{c}{N_{1}^{\lambda}(t)} \\ {Z_{1}^{\lambda}(t)}\end{array}\right)&=\int d^{3} r\;g(x')\left(\begin{array}{c} {\dot{x}'\rho _{n}^{\lambda } (\vec{r},t)+\hat{e}\cdot \vec{j}_{n}^{\lambda } (\vec{r},t)} \\ {\dot{x}'\rho _{p}^{\lambda } (\vec{r},t)+\hat{e}\cdot \vec{j}_{p}^{\lambda } (\vec{r},t)} \end{array}\right)\nonumber\\
&=\left(\begin{array}{c} {v_{n}^{\lambda } (t)} \\ {v_{p}^{\lambda } (t)} \end{array}\right).
\end{align}
In this expression and below, for convenience, we replace the delta function $\delta(x)$ by a
Gaussian $g(x)=\left(1/\kappa\sqrt{2\pi}\right)\exp\left(-x^{2}/2\kappa^{2}\right)$ which behaves
almost like delta function for sufficiently small $\kappa$. In the numerical calculations dispersion
of the Gaussian is taken in the order of the lattice side $\kappa=1.0$~fm. The right side of Eq.
\eqref{eq8} defines the drift coefficients $v_{\alpha}^{\lambda}(t)$ for the neutrons and the
protons for the event $\lambda$. In the SMF approach the current density vector is given by,
\begin{align} \label{eq9}
\vec{j}_{\alpha}^{\lambda}(\vec{r},t)&=\frac{\hbar}{2im}\sum_{ij\in\alpha}\left[\Phi_{j}^{*\alpha}(\vec{r},t;\lambda)\vec{\nabla}\Phi_{i}^{\alpha}(\vec{r},t;\lambda)\right.\nonumber\\
&\qquad\qquad\quad\left.-\Phi_{i}^{\alpha}(\vec{r},t;\lambda)\vec{\nabla}\Phi_{j}^{*\alpha}(\vec{r},t;\lambda)\right]\rho _{ji}^{\lambda}\nonumber\\
&=\frac{\hbar}{m}\sum _{ij\in\alpha}\text{Im}\!\left(\Phi_{j}^{*\alpha}(\vec{r},t;\lambda)\vec{\nabla}\Phi_{i}^{\alpha}(\vec{r},t;\lambda)\rho_{ji}^{\lambda}\right).
\end{align}
Equation~\eqref{eq8} provides a Langevin description for the stochastic evolution the neutron and the
proton numbers of the projectile-like fragments. Drift coefficients $v_{\alpha}^{\lambda}(t)$
fluctuate from event to event due to stochastic elements of the initial density matrix
$\rho_{ji}^{\lambda}$ and also due to the different sets of the wave functions in different events.
As a result, there are two sources for fluctuations of the nucleon drift coefficients: (i)
fluctuations those arise from the different set of single-particle wave functions in each event, and
(ii) the explicit fluctuations $\delta v_{p}^{\lambda}(t)$ and $\delta v_{n}^{\lambda}(t)$ arising
from the stochastic part of proton and neutron currents.

\subsection{Mean neutron and proton drift path}
Equations for the mean values of proton $Z_{1} (t)=\overline{Z_{1}^{\lambda}(t)}$ and neutron
$N_{1}(t)=\overline{N_{1}^{\lambda}(t)}$ numbers of the projectile-like fragments are obtained by
taking the ensemble averaging of the Langevin equation~\eqref{eq8}.  For small amplitude fluctuations, and
using the fact that average values of density matrix elements are given by the average occupation
numbers as $\overline{\rho_{ji}^{\lambda}}=\delta_{ji}n_{j}$, we obtain the usual mean-field result
given by the TDHF equations,
\begin{align} \label{eq10}
\frac{d}{dt}\left(\begin{array}{c} {N_{1} (t)} \\ {Z_{1} (t)} \end{array}\right)&=\int d^{3} r\;g(x')\left(\begin{array}{c} {\dot{x}'\rho _{n} (\vec{r},t)+\hat{e}\cdot \vec{j}_{n} (\vec{r},t)} \\ {\dot{x}'\rho _{p} (\vec{r},t)+\hat{e}\cdot \vec{j}_{p} (\vec{r},t)} \end{array}\right)\nonumber\\
&=\left(\begin{array}{c} {v_{n} (t)} \\ {v_{p} (t)} \end{array}\right).
\end{align}
Here, the mean values of the densities and the currents densities of neutron and protons are given by,
\begin{align} \label{eq11}
\rho_{\alpha}(\vec{r},t)=\sum_{h\in\alpha}\Phi_{h}^{*\alpha}(\vec{r},t)\Phi_{h}^{\alpha}(\vec{r},t)
\end{align}
and
\begin{align} \label{eq12}
\vec{j}_{\alpha}(\vec{r},t)=\frac{\hbar}{m}\sum_{h\in\alpha}\text{Im}\!\left(\Phi_{h}^{*\alpha}(\vec{r},t)\vec{\nabla}\Phi_{h}^{\alpha}(\vec{r},t)\right),
\end{align}
where the summation $h$ runs over the occupied states originating both from the projectile and the
target nuclei. The drift coefficients $v_{p}(t)$ and $v_{n}(t)$ denote the net proton and neutron
currents across the window.

\begin{figure}[!hpt]
%\vspace{0.2cm}
\includegraphics*[width=4.5cm]{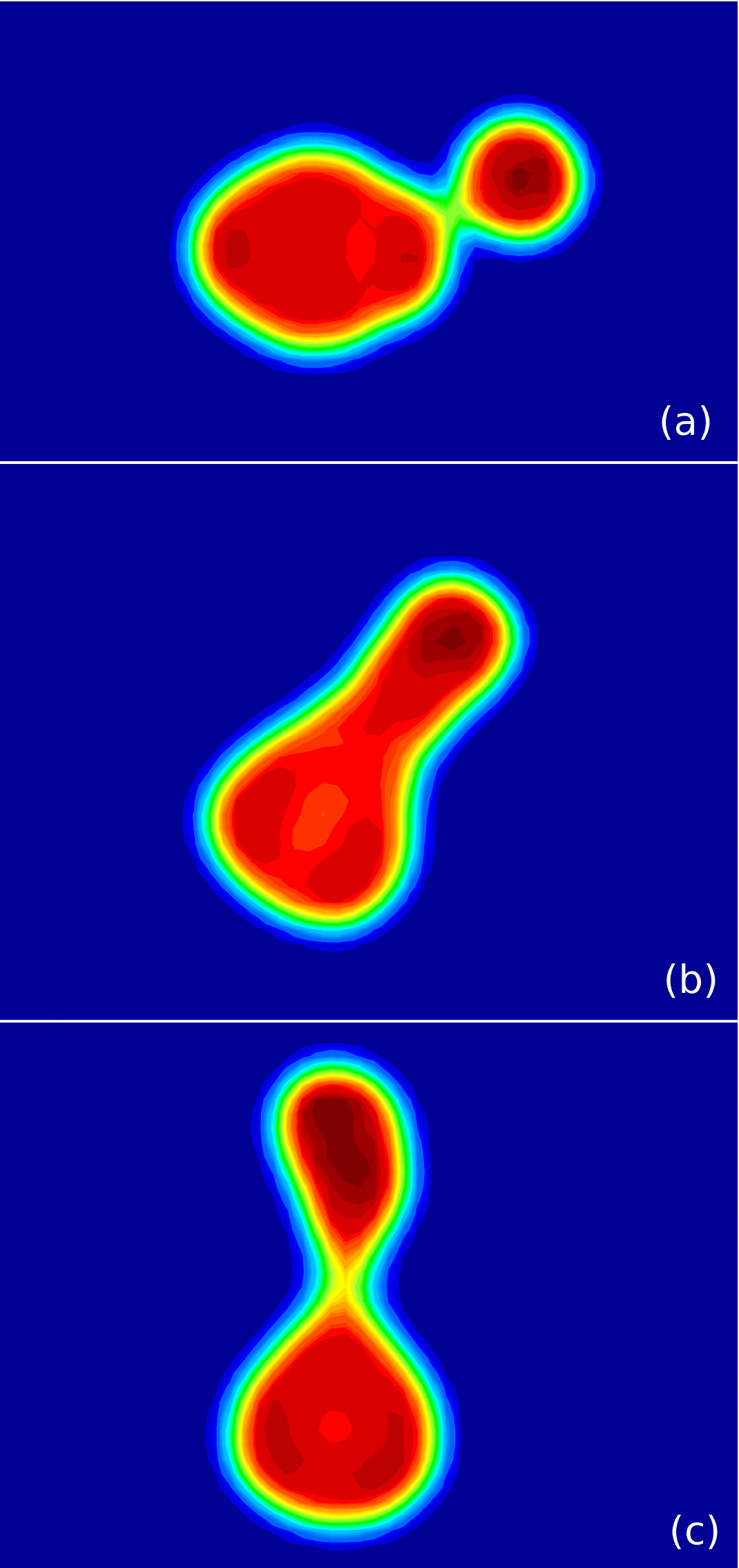}
%\vspace{0.2cm}
\caption{(color online) The density profile of the $^{48}\mathrm{Ca}+{}^{238}\mathrm{U}$ collisions in the tip geometry of the uranium nucleus with the bombarding energy $E_{\text{c.m.}}=193$~MeV at an impact parameter $b=2.8$~fm or equivalently at the initial orbital angular momentum $\ell=54\,\hbar$ at times $t=300$~fm/c (a), $t=1170$~fm/c (b) and $t=2070$~fm/c (c).}
\label{fig1}
\end{figure}

\begin{figure}[!hpt]
%\vspace{0.2cm}
\includegraphics*[width=8.5cm]{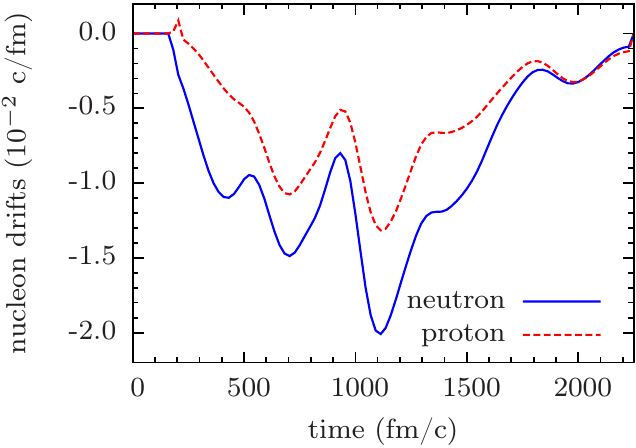}
%\vspace{0.2cm}
\caption{(color online) The neutron and proton drift coefficient the $^{48}\mathrm{Ca}+{}^{238}\mathrm{U}$ collisions in the tip geometry of the uranium nucleus with the bombarding energy $E_{\text{c.m.}}=193$~MeV at an impact parameter $b=2.8$~fm or equivalently at the initial orbital angular momentum $l=54\,\hbar$.}
\label{fig2}
\end{figure}

\begin{figure}[!hpt]
%\vspace{0.2cm}
\includegraphics*[width=8.5cm]{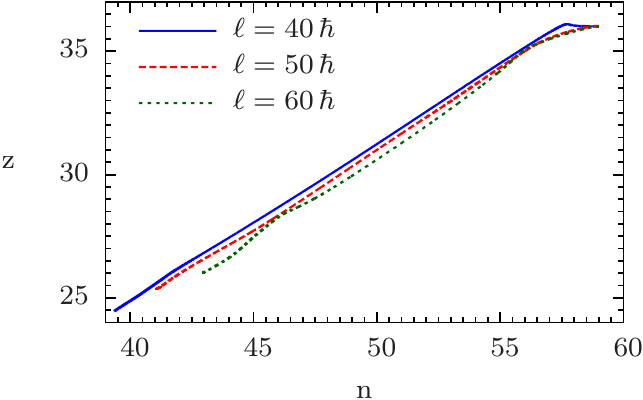}
%\vspace{0.2cm}
\caption{(color online) The neutron and proton mean-drift path in the $(N,Z)$ plane, in the $^{48}\mathrm{Ca}+{}^{238}\mathrm{U}$ collisions in the tip geometry of the uranium with the bombarding energy $E_{\text{c.m.}}=193$~MeV at the impact parameters $b=2.1$~fm, $b=2.6$~fm and $b=3.1$~fm or equivalently at the initial angular momentum $\ell=40\,\hbar$, $\ell=50\,\hbar$ and $\ell=60\,\hbar$. Here $n=N_{0}-N_{1}$ and $z=Z_{0}-Z_{1}$, with $N_{1}$ and $Z_{1}$ indicating the light fragments.}
\label{fig3}
\end{figure}

Since the uranium nucleus has large quadrupole deformation, the collision takes place in many
different geometries. In this work, we observe that the dominant contribution to the fragment
distributions in the $^{48}\mathrm{Ca}+{}^{238}\mathrm{U}$ collisions at the bombarding energy $E_{\text{c.m.}}=193$
MeV, reported in~\cite{kozulin2014b}, are coming from the tip geometry of the uranium nucleus.
Therefore in analyzing the measured data we incorporate only collisions involving the tip configuration of
the uranium. Figure~\ref{fig1} illustrate the density profile of the $^{48}\mathrm{Ca}+{}^{238}\mathrm{U}$ collisions
in the tip geometry of the uranium with the bombarding energy $E_{\text{c.m.}} =193$~MeV at an
impact parameter $b=2.8$~fm or equivalently at the initial orbital angular momentum $\ell=54\hbar$
at several times during the collision. This computation and all other numerical computation in this
work are carried out by employing 3D THDF program of Umar et al.~\cite{umar1991a,umar2006c}. The SLy4d Skyrme
interaction~\cite{chabanat1998a,kim1997} is used.
Figure~\ref{fig2} illustrate neutron and proton drift coefficients at the same energy and the same impact
parameter. The fluctuations of the drift coefficients as function of time is a result of the shell
structure of the population of different nuclei during the evolution of the projectile-like and
target-like nuclei. Even the more interesting presentation of the results of the mean-field
evolution in the tip geometry is presented in Fig.~\ref{fig3} at the same bombarding energy for a
few different impact parameters. This figure shows the mean drift-path of di-nuclear system in the
$(N,Z)$ plane. After touching, we observe a rapid charge equilibration, which is not very visible
due to the fact that the colliding nuclei and the composite system have nearly same charge
asymmetry. After touching, di-nuclear system drift toward symmetry along the valley of the beta
stability line nearly the similar manner in collisions for different impact parameters. The neutron
and proton numbers of the symmetric equilibrium state are $N_{0}=\left(N_{P}+N_{T}\right)/2=87$ and
$Z_{0}=\left(Z_{P}+Z_{T}\right)/2=56$. As seen from the mean drift path, during the mean evolution,
the system separate before reaching the symmetric equilibrium state.

\subsection{Co-variances of  fragment charge and mass distributions}
Our task is to evaluate the fluctuations of the neutron and proton numbers around their mean values.
For this purpose, we linearize the Langevin Eq.~\eqref{eq8} around their mean values $Z_{1}(t)$ and
$N_{1}(t)$. There are two different sources of the fluctuations. First contribution arises from the
different set of wave functions in different events. In the leading order, we can express this
effect as deviations of the drift coefficients from their mean values in terms of fluctuations in
neutron and proton numbers. The second contribution arises from the initial fluctuations of the
elements of the density matrix. As a result the linearized form of the Langevin Eq.~\eqref{eq8}
becomes,
\begin{align} \label{eq13}
\frac{d}{dt}\left(\begin{array}{c}{\delta Z_{1}(t)} \\ {\delta N_{1}(t)} \end{array}\right)=&\left(\begin{array}{c} {\frac{\partial v_{p}}{\partial Z_{1}}\left(Z_{1}^{\lambda}-Z_{1}\right)+\frac{\partial v_{p}}{\partial N_{1}} \left(N_{1}^{\lambda}-N_{1} \right)} \\ {\frac{\partial v_{n} }{\partial Z_{1} } \left(Z_{1}^{\lambda } -Z_{1} \right)+\frac{\partial v_{n} }{\partial N_{1} } \left(N_{1}^{\lambda } -N_{1} \right)} \end{array}\right)\nonumber\\
&+\left(\begin{array}{c} {\delta v_{p}^{\lambda } (t)} \\ {\delta v_{n}^{\lambda } (t)} \end{array}\right).
\end{align}
The linear limit provides a good approximation for small amplitude fluctuations and it becomes even
better if the fluctuations are nearly harmonic around the mean values. The derivatives of drift
coefficients are evaluated on the mean trajectory and the quantities $\delta
v_{\alpha}^{\lambda}(t)$ denote the stochastic part of drift coefficients given by,
\begin{align} \label{eq14}
\delta v_{\alpha}^{\lambda}(t)=\int d^{3} r\,g(x')\left(\dot{x}'\delta\rho_{\alpha}^{\lambda}(\vec{r},t)+\hat{e}\cdot\delta\vec{j}_{\alpha}^{\lambda}(\vec{r},t)\right),
\end{align}
with the fluctuating neutron and proton current densities
\begin{align} \label{eq15}
\delta \vec{j}_{\alpha}^{\lambda}(\vec{r},t)=\frac{\hbar}{m}\sum_{ij\in\alpha}\text{Im}\!\left(\Phi_{j}^{*\alpha}(\vec{r},t)\vec{\nabla}\Phi_{i}^{\alpha}(\vec{r},t)\delta\rho_{ji}^{\lambda}\right),
\end{align}
and the fluctuating neutron and proton number densities
\begin{align} \label{eq16}
\delta\rho_{\alpha}^{\lambda}(\vec{r},t)=\sum_{ij\in\alpha}\Phi_{j}^{*\alpha}(\vec{r},t)\delta\rho_{ji}^{\lambda}\Phi_{i}^{\alpha}(\vec{r},t).
\end{align}
The variances and the co-variance of neutron and proton distribution of projectile fragments are
defined as $\sigma_{NN}^{2}(t)=\overline{\left(N_{1}^{\lambda}-N_{1}\right)^{2}}$,
$\sigma_{ZZ}^{2}(t)=\overline{\left(Z_{1}^{\lambda}-Z_{1}\right)^{2}}$, and
$\sigma_{NZ}^{2}(t)=\overline{\left(N_{1}^{\lambda}-N_{1}\right)\left(Z_{1}^{\lambda}-Z_{1}\right)}$.
Multiplying both side of Langevin equations \eqref{eq13} by $N_{1}^{\lambda}-N_{1}$ and
$Z_{1}^{\lambda}-Z_{1}$, and taking the ensemble average, we find evolution of the co-variances are
specified by the following set of coupled differential equations
\cite{schroder1981,merchant1981},
\begin{align} \label{eq17}
\frac{\partial }{\partial t}\sigma_{NN}^{2}=2\frac{\partial v_{n}}{\partial N_{1}}\sigma_{NN}^{2}+2\frac{\partial v_{n}}{\partial Z_{1}}\sigma_{NZ}^{2}+2D_{NN},
\end{align}
\begin{align} \label{eq18}
\frac{\partial }{\partial t}\sigma_{ZZ}^{2}=2\frac{\partial v_{p}}{\partial Z_{1}}\sigma_{ZZ}^{2}+2\frac{\partial v_{p}}{\partial N_{1}}\sigma_{NZ}^{2}+2D_{ZZ},
\end{align}
and
\begin{align} \label{eq19}
\frac{\partial}{\partial t}\sigma_{NZ}^{2}=\frac{\partial v_{p}}{\partial N_{1}}\sigma_{NN}^{2}+\frac{\partial v_{n}}{\partial Z_{1}}\sigma_{ZZ}^{2}+\sigma_{NZ}^{2}\left(\frac{\partial v_{p}}{\partial Z_{1}}+\frac{\partial v_{n}}{\partial N_{1}}\right).
\end{align}
In these expressions, $D_{NN}$ and $D_{ZZ}$ denote the neutron and proton quantal diffusion
coefficients which are discussed in the next section. It is well know that the Langevin Eq.
\eqref{eq13} is equivalent to the Fokker-Planck equation for the distribution function
$P_{b}(N,Z,t)$ of the macroscopic variables~\cite{risken1996}. In the tip geometry there is a
cylindrical symmetry for the distribution function for each impact parameter. As a result, for each
impact parameter $b$ (or the initial orbital angular momentum $\ell$), the proton and neutron
distribution function $P_{b}(N,Z,t)$ of the project-like or the target-like fragments is a
correlated Gaussian function described by the mean values and the co-variances as,
\begin{align} \label{eq20}
P_{b}(N,Z,t)=\frac{1}{2\pi\sigma_{NN}(b)\sigma_{ZZ}(b)\sqrt{1-\rho_{b}^{2}}}\exp\left(-C_{b}\right).
\end{align}
Here, the argument of the exponent $C_{b}$ for each impact parameter is given by
\begin{align} \label{eq21}
C_{b}=\frac{1}{2\left(1-\rho_{b}^{2}\right)}&\left[\left(\frac{Z-Z_{b}}{\sigma_{ZZ}(b)}\right)^{2}-2\rho\left(\frac{Z-Z_{b}}{\sigma_{ZZ}(b)}\right)\left(\frac{N-N_{b}}{\sigma_{NN}(b)}\right)\right.\nonumber\\
                                            &\left.\;\,+\left(\frac{N-N_{b}}{\sigma_{NN}(b)}\right)^{2}\right],
\end{align}
with $\rho_{b}=\sigma_{NZ}^{2}(b)/\sigma_{ZZ}(b)\sigma_{NN}(b)$. The mean values $N_{b}=\overline{N}_{b}$, $Z_{b}=\overline{Z}_{b}$ denote the mean neutron and proton numbers of the target-like or project-like fragments. The set of coupled Eqs.~(\ref{eq17}-\ref{eq19}) for co-variances are familiar from the phenomenological nucleon exchange model, and they were derived from the Fokker-Planck equation for the fragment neutron and proton distributions in the deep-inelastic heavy-ion collisions~\cite{schroder1981,merchant1981}.
Equation~\eqref{eq20} determines the joint probability distribution at a given impact parameter. Then, it is possible to calculate the cross-section $\sigma(N,Z)$ for production of nuclei with neutron and proton numbers by integrating the probability distributions over the range of the impact parameters corresponds to the experimental data as,
\begin{align} \label{eq22}
\sigma(N,Z)=\int_{b_{1}}^{b_{2}}2\pi b\,P_{b}(N,Z)\,db,
\end{align}
where $P_{b}(N,Z)$ denotes the distribution functions at the separation instant of the fragments.
Distribution $P_{b}(A-A_b)$ of the mass number of the fragments is obtained by substituting $Z=A-N$ in
$P_{b}(N,Z)$ and integrating over $N$. For a given impact parameter this yields a Gaussian function
for the mass number distribution of the target-like or projectile-like fragments,
\begin{align} \label{eq23}
P_{b}(A-A_b)=\frac{1}{\sqrt{2\pi}\sigma_{AA}(b)}\exp\left[-\frac{1}{2}\left(\frac{A-A_{b}}{\sigma_{AA}(b)}\right)^{2}\right],
\end{align}
where $A_{b}=N_{b}+Z_{b}$ is the mean value of the mass number and the variance given by
$\sigma_{AA}^{2}(b)=\sigma_{NN}^{2}(b)+\sigma_{ZZ}^{2}(b)+2\sigma_{NZ}^{2}(b)$. The cross-section
for production of nuclei with a mass number $A=N+Z$ is calculated in the similar manner to Eq.
\eqref{eq22}.

\section{Transport Coefficients}
\label{transport}
\subsection{Quantal Diffusion Coefficients}
Stochastic part of the drift coefficients $\delta v_{p}^{\lambda}(t)$ and $\delta v_{n}^{\lambda}(t)$ are specified by uncorrelated Gaussian distributions. Stochastic drift coefficients have zero mean values $\overline{\delta v_{p}^{\lambda}(t)}=0$, $\overline{\delta v_{n}^{\lambda}(t)}=0$ and the associated correlation functions~\cite{gardiner1991,weiss1999},
\begin{align} \label{eq24}
\int_{0}^{t}dt'\overline{\delta v_{\alpha }^{\lambda } (t)\delta v_{\alpha }^{\lambda } (t')} =D_{\alpha \alpha } (t),
\end{align}
determine the diffusion coefficients $D_{\alpha\alpha}(t)$ for proton and neutron transfers. As seen
from Eq.~\eqref{eq14}, there are two different contribution to the stochastic part of the drift
coefficients: (i) density fluctuations in vicinity of the rotating window plane which involves
collective velocity of the window and (ii) current density fluctuation across the rotating window.
Since nucleon flow velocity through the window is much larger than the collective velocity of the
window, the current density fluctuations current make the dominant contribution. Therefore in our
analysis, we retain only the current density fluctuations in the stochastic part of the drift
coefficients,
\begin{align} \label{eq25}
\delta v_{\alpha}^{\lambda}(t)=\frac{\hbar}{m}\int & d^{3} r\,g(x')\nonumber\\
 &\times\sum_{ij\in\alpha}\text{Im}\left(\Phi_{j}^{*\alpha}(\vec{r},t)\hat{e}\cdot\vec{\nabla}\Phi_{i}^{\alpha}(\vec{r},t)\delta\rho_{ji}^{\lambda}\right).
\end{align}
Furthermore, in the stochastic part of the drift coefficients, we impose a physical constraint on
the summations of single-particle states. The transitions among the single particle states
originating from the projectile or target nuclei do not contribute to nucleon exchange mechanism.
Therefore in Eq.~\eqref{eq25}, we restrict summation as follows: when the summation $i\in T$ runs
over the states originating from the target nucleus, the summation $j\in P$ runs over the states
originating from the projectile, and vice versa. Using the main postulate of the SMF approach given
by Eq.~\eqref{eq4}, we can calculate the correlation functions of the stochastic part of the drift
coefficients. At zero temperature, since the average occupation factor are zero or one, we find the
correlation functions is expressed as,
\begin{align} \label{eq26}
\overline{\delta v_{\alpha}^{\lambda}(t)\delta v_{\alpha}^{\lambda}(t')}=\text{Re}&\left(\sum_{p\in P,h\in T}A_{ph}^{\alpha}(t)A_{ph}^{*\alpha}(t')\right.\nonumber\\
&\left.\quad+\sum_{p\in T,h\in P}A_{ph}^{\alpha}(t)A_{ph}^{*\alpha}(t')\right).
\end{align}
Here the matrix elements are given by
\begin{align} \label{eq27}
A_{ph}^{\alpha}(t)=\frac{\hbar}{2m}\int d^{3} r\,g(x')&\left(\Phi_{p}^{*\alpha}(\vec{r},t)\hat{e}\cdot\vec{\nabla}\Phi_{h}^{\alpha}(\vec{r},t)\right.\nonumber\\
&\left.\quad-\Phi_{h}^{\alpha}(\vec{r},t)\hat{e}\cdot\vec{\nabla}\Phi_{p}^{*\alpha}(\vec{r},t)\right).
\end{align}
We note that employing a partial integration, we can put this expression in the following form,
\begin{align} \label{eq28}
A_{ph}^{\alpha}(t)=\frac{\hbar}{m}\int & d^{3} r\,g(x')\Phi_{p}^{*\alpha}(\vec{r},t)\nonumber\\
&\quad\times\left(\hat{e}\cdot\vec{\nabla}\Phi_{h}^{\alpha}(\vec{r},t)-\frac{x'}{2\kappa^{2}}\Phi_{h}^{\alpha}(\vec{r},t)\right).
\end{align}
In order to evaluate the correlation function Eq.~\eqref{eq26} of the stochastic drift coefficient,
we introduce the following approximate treatment. In the first term of the right hand side of Eq.
\eqref{eq26}, we add and subtract the hole contributions to give,
\begin{align} \label{eq29}
\sum_{p\in P,h\in T}A_{ph}^{\alpha}(t)A_{ph}^{*\alpha}(t')=&\sum_{a\in P,h\in T}A_{ah}^{\alpha}(t)A_{ah}^{*\alpha}(t')\nonumber\\
&-\sum_{h'\in P,h\in T}A_{h'h}^{\alpha}(t)A_{h'h}^{*\alpha}(t').
\end{align}
Here, the summation $a$ run over the complete set of states originating from the projectile. In the
first term, we cannot use the closure relation to eliminate the complete set of single-particle
states, because the wave functions are evaluated at different times. However, we note that the
time-dependent single-particle wave functions during short time intervals exhibit nearly a diabatic
behavior~\cite{norenberg1981}. In another way of stating that during short time intervals the nodal
structure of time-dependent wave functions do not change appreciably. Most dramatic diabatic
behavior of the time-dependent wave-functions is apparent in the fission dynamics. The TDHF
solutions force the system to follow the diabatic path, which prevents the system to break up into
fragments. As a result of these observations, we introduce, during short time $\tau =t-t'$ evolutions,
in the order of the correlation time, a diabatic approximation into the time dependent
wave-functions by shifting the time backward (or forward) according to,
\begin{align} \label{eq30}
\Phi_{a}(\vec{r},t')\approx\Phi_{a}(\vec{r}-\vec{u}\tau,t).
\end{align}
where $\vec{u}$ denotes a suitable flow velocity of nucleons through the window. Now, we can employ
the closure relation to obtain,
\begin{align} \label{eq31}
\sum_{a}\Phi_{a}^{*}(\vec{r}_{1},t)\Phi_{a}(\vec{r}_{2}-\vec{u}\tau,t)=\delta(\vec{r}_{1}-\vec{r}_{2}+\vec{u}\tau),
\end{align}
where, the summation $a$ runs over the complete set of states originating from target or projectile,
and the closure relation is valid for each set of the spin-isospin degrees of freedom. The flow
velocity $\vec{u}(\vec{R},T)$ may depend on the mean position $\vec{R}=(\vec{r}_{1}+\vec{r}_{2})/2$
and the time mean $T=(t+t')/2$ . Employing the closure relation in the first term of the right hand
side of Eq.~\eqref{eq29}, we find
\begin{align} \label{eq32}
\sum_{a\in P,h\in T}A_{ah}^{\alpha}(t)A_{ah}^{*\alpha}(t')=\sum_{h\in T}\int & d^{3} r_{1}d^{3}r_{2}\delta(\vec{r}_{1}-\vec{r}_{2}+\vec{u}_{h}\tau)\nonumber\\
&\times W_{h}^{\alpha}(\vec{r}_{1},t)W_{h}^{*\alpha}(\vec{r}_{2},t').
\end{align}
The closure relation in Eq.~\eqref{eq31} is valid for any choice of the flow velocity. The most
suitable choice is the flow velocity of the hole state $\vec{u}_{h}(\vec{R},T)$ in each term in the
summation, which is taken in this expression. In this manner the complete set of single-particle
states is eliminated and the calculations of the quantal diffusion coefficients are greatly
simplified. In fact, in order to calculate this expression, we only need the hole states originating
from target which are provided by the TDHF description. The local flow velocity of each
wave-function is specified by the usual expression of the current density divided by the particle
density as given in Eq.~(B8) in Appendix B. The quantity $W_{h}^{\alpha}(\vec{r}_{1},t)$ is given
by,
\begin{align} \label{eq33}
W_{h}^{\alpha } (\vec{r}_{1} ,t)=&\frac{\hbar }{m} g(x'_{1} )\nonumber\\
&\times\left(\hat{e}\cdot \vec{\nabla }_{1} \Phi _{h}^{\alpha } (\vec{r}_{1} ,t)-\frac{x'_{1} }{2\kappa ^{2} } \Phi _{h}^{\alpha } (\vec{r}_{1} ,t)\right),
\end{align}
and $W_{h}^{*\alpha}(\vec{r}_{2},t')$ is given by a similar expression. A detailed analysis of Eq.
\eqref{eq32} is presented in Appendix B. The result of this analysis as given by Eq.~(B19) is,
\begin{align} \label{eq34}
\sum _{a\in P,h\in T}A_{ah}^{\alpha } (t)A_{ah}^{*\alpha } (t') =\int & d^{3} r \tilde{g}(x')G_{T} (\tau )\nonumber\\
&\times J_{\bot ,\alpha }^{T} (\vec{r},t-\tau /2).
\end{align}
Here $J_{\bot ,\alpha }^{T} (\vec{r},t-\tau /2)$ represents the sum of the magnitude of current
densities perpendicular to the window due to the each wave functions originating from target,
\begin{align} \label{eq35}
J_{\bot ,\alpha }^{T} (\vec{r},t-\tau /2)=\frac{\hbar }{m} \sum _{h\in T}&\left|\frac{}{}\text{Im}\left[\Phi _{h}^{*\alpha } (\vec{r},t-\tau /2)\right.\right.\nonumber\\
&\left.\left.\times\left(\hat{e}\cdot \vec{\nabla }\Phi _{h}^{\alpha} (\vec{r},t-\tau /2)\right)\right]\right|.
\end{align}
The quantity $G_{T}(\tau)$ is given by Eq.~(B20), and it is the average value of the memory kernels
$G_{T}^{h}(\tau)$ of Eq.~(B13). It is possible to carry out a similar analysis in the second term in
the right side of Eq.~\eqref{eq13} to give,
\begin{align} \label{eq36}
\sum _{a\in T,h\in P}A_{ah}^{\alpha}(t)A_{ah}^{*\alpha}(t')=\int & d^{3} r\,\tilde{g}(x')G_{P} (\tau )\nonumber\\
&\times J_{\bot ,\alpha }^{P} (\vec{r},t-\tau /2).
\end{align}
In a similar manner, $J_{\bot,\alpha}^{P}(\vec{r},t-\tau /2)$ is determined by the sum of the
magnitude of the current densities due wave functions originating from projectile. In Eq.
\eqref{eq35} and Eq.~\eqref{eq36} we use lover case $\vec{r}$ instead of capital letter. As a
result, the quantal expressions of the proton and neutron diffusion coefficients takes the form,
\begin{align} \label{eq37}
D_{\alpha\alpha}(t)= & \int_{0}^{t}d\tau\int d^{3} r\,\tilde{g}(x')\left[G_{T}(\tau)J_{\bot,\alpha}^{T}(\vec{r},t-\tau/2)\right.\nonumber\\
 &\qquad\qquad\qquad\quad\left.+G_{P}(\tau)J_{\bot ,\alpha }^{P} (\vec{r},t-\tau /2)\right]\nonumber\\
&-\int_{0}^{t}d\tau\,\text{Re}\!\left(\sum_{h'\in P,h\in T}A_{h'h}^{\alpha}(t)A_{h'h}^{*\alpha}(t-\tau)\right.\nonumber\\
&\qquad\qquad\quad\left.+\sum_{h'\in T,h\in P}A_{h'h}^{\alpha}(t)A_{h'h}^{*\alpha}(t-\tau)\right).
\end{align}
These quantal expressions for the nucleon diffusion coefficients for heavy-ion collisions at finite
impact parameters provide an extension of the result published in~\cite{ayik2017} for the central
collisions. In general, we observe that there is a close analogy between the quantal expression and
the classical diffusion coefficient in a random walk problem
\cite{gardiner1991,weiss1999,randrup1979}. The first term in the quantal expression gives the sum
of the nucleon currents across the window from the target-like fragment to the projectile-like
fragment and from the projectile-like fragment to the target-like fragment,which is integrated over
the memory. This is analogous to the random walk problem, in which the diffusion coefficient is
given by the sum of the rate for the forward and backward steps. The second term in the quantal
diffusion expression stands for the Pauli blocking effects in nucleon transfer mechanism, which does
not have a classical counterpart. It is important to note that the quantal diffusion coefficients
are entirely determined in terms of the occupied single-particle wave functions obtained from the
TDHF solutions.

\begin{figure}[!hpt]
%\vspace{0.2cm}
\includegraphics*[width=8.5cm]{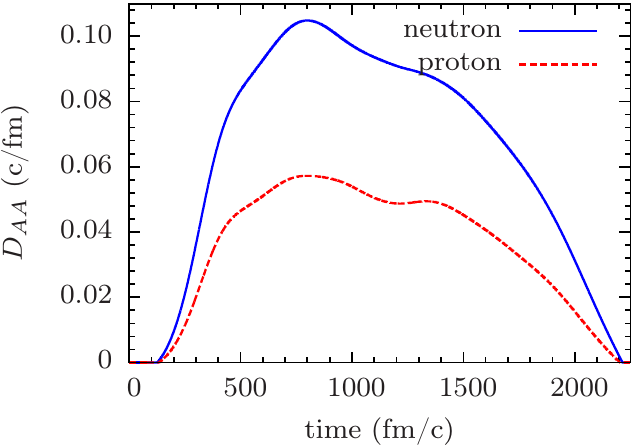}
%\vspace{0.2cm}
\caption{(color online) The neutron and proton diffusion coefficients as a function of time in the
$^{48}\mathrm{Ca}+{}^{238}\mathrm{U}$ collisions in the tip geometry of the uranium with the bombarding energy
$E_{\text{c.m.}} =193$~MeV at the impact parameter $b=2.8$~fm or equivalently at the initial orbital
angular momentum $\ell=54\,\hbar$.}
\label{fig4}
\end{figure}

From our calculations, we find that the average nucleon flow speed across the window between the
colliding nuclei is around $u_{\bot}\approx 0.05c$. Using the expression $\tau=\kappa/|u_{\bot}|$,
given below Eq.~(B20) for nucleon flows from target-like or projectile-like fragments, with a
dispersion $\kappa =1.0$~fm, we find the average memory time is around $\tau\approx 20$~fm/c. In the
nuclear one-body dissipation mechanism, it is possible to estimate the memory time in terms of a
typical nuclear radius and the Fermi speed as $\tau_{0}\approx R/v_{F}$. If we take $R\approx 5.0$
fm and $v_{F}\approx 0.2c$, it gives the same order of magnitude for the memory time
$\tau_{0}\approx 25$~fm/c. Since this time interval is much shorter than typical interaction time of collisions
($\tau_{0} << 1000$~fm/c), we find that the memory effect is not very significant in nucleon exchange
mechanism. The time integration of the average memory kernel for the nucleon transfers from the
target-like fragments becomes,
\begin{align} \label{eq38}
\tilde{G}_{T}(t)&=\int_{0}^{t}ds G_{T}(s)\nonumber\\
&=\int_{0}^{t}ds\frac{1}{\sqrt{4\pi}}\frac{1}{\tau}e^{-(s/2\tau)^{2}}=\frac{1}{2}\text{erf}(t/2\tau ).
\end{align}
A similar expression for the nucleon transfers from the projectile-like fragments is given by
$\tilde{G}_{P} (t)=\frac{1}{2}\text{erf}(t/2\tau)$. For $t>>\tau$ the error function behaves like a
delta function, $\text{erf}(t/2\tau)\to\delta(t)$. Therefore we can neglect the memory effect in the
first line of the diffusion coefficient.  For the same reason, memory effect is not very
effective in the Pauli blocking terms in Eq.~\eqref{eq37} as well, however in the calculations we
keep the memory integrals in these terms. As an example, Fig.~\ref{fig4} shows the neutron and
proton diffusion coefficients as a function of time in the $^{48}\mathrm{Ca}+{}^{238}\mathrm{U}$ collisions in the
tip geometry of the uranium with the bombarding energy $E_{\text{c.m.}} =193$~MeV at the impact
parameter $b=2.8$~fm or equivalently at the initial orbital angular momentum $\ell=54\,\hbar$. As
seen from the figure, neutron diffusion coefficient is nearly a factor of two larger than the proton
diffusion coefficient.

\subsection{Derivatives of drift coefficients}
In order to determine the co-variances from Eqs.~(\ref{eq17}-\ref{eq19}), in addition to the
diffusion coefficients $D_{ZZ}$ and $D_{NN}$, we need to know the rate of change of drift
coefficients in the vicinity of their mean values. In order to calculate rates of change of the
drift coefficients, we should calculate neighboring events in the vicinity of the mean-field path.
Here, instead of such a detailed description, in order to determine the derivatives of the drift
coefficients, we employ the fluctuation-dissipation theorem, which provides a general relation
between the diffusion and drift coefficients in the transport mechanism of the relevant collective
variables as often used in the phenomenological approaches~\cite{randrup1979,merchant1982}. Proton and neutron
diffusions in the N-Z plane are driven in a correlated manner by the potential energy surface of the
di-nuclear system. As a consequence of the symmetry energy, the diffusion in the direction
perpendicular to the mean-drift path (the beta stability valley) takes place rather rapidly leading
to a fast equilibration of the charge asymmetry, and the diffusion continues rather slowly along the
beta-stability valley. In Fig.~\ref{fig3}, the calculations carried out by the TDHF equations
illustrate very nicely the expected the mean-drift paths in the collision of $^{48}\mathrm{Ca}+{}^{238}\mathrm{U}$
system with the several impact parameters. Since the charge asymmetries of $^{48}$Ca and $^{238}$U
are very close to the charge asymmetry of the composite system, a rapid equilibration of the charge
asymmetry is not very visible in this system. We observe that the di-nuclear system drifts towards
symmetry during long contact time, but separates before reaching to the symmetry. Following this
observation and borrowing an idea from references~\cite{merchant1981,merchant1982}, for each impact
parameter, we parameterize the $N_{1}$ and $Z_{1}$ dependence of the potential energy surface of the
di-nuclear system in terms of two parabolic forms,
\begin{align} \label{eq39}
U(N_{1},Z_{1})=&\frac{1}{2}\alpha\left(\tilde{z}\cos\phi-\tilde{n}\sin\phi\right)^{2}\nonumber\\
&+\frac{1}{2}\beta\left(z\sin\phi+n\cos\phi\right)^{2}.
\end{align}
Here, the first term describes a strong driving force perpendicular to the mean-drift path. The
quantity $\tilde{z}\cos\phi-\tilde{n}\sin\phi$ denotes the perpendicular distance of a di-nuclear
state with $N_{1}$, $Z_{1}$ from the mean-drift with $\tilde{z}=Z_{f}-Z_{1}$ and
$\tilde{n}=N_{f}-N_{1}$. The angle between the mean-drift path and the $N$ axis is indicated by
$\phi$. Here $N_{f}$ and $Z_{f}$ are the mean values of neutron and proton numbers of the light
fragment just after the separation.  The second parabola describes a relative weak driving force
toward symmetry along the stability valley. The quantity $z\sin\phi+n\cos\phi$ indicates the
distance of the di-nuclear state with $N_{1}$, $Z_{1}$ state along the mean-drift path from the
symmetry with $z=Z_{0}-Z_{1}$ and $n=N_{0}-Z_{1}$. The quantities $N_{0}$ and $Z_{0}$ denotes the
equilibrium values of the neutron and proton numbers, which are determined by the average values of
the neutron and proton numbers of the projectile and target, $N_{0} =\left(N_{P} +N_{T} \right)/2$
and $Z_{0} =\left(Z_{P} +Z_{T} \right)/2$. The parameters of the driving potential depend on the
impact parameter. We can determine the values of $Z_{f}$ and $N_{f}$ from the mean-drift path for
each impact parameter. Also, we observe from Fig.~\ref{fig3} that the slope of the mean-drift paths
is nearly the same for different impact parameters. Therefore the angle $\phi \approx 32^\circ$ is
approximately the same for different impact parameters. Following from the fluctuation-dissipation
theorem, it is possible to relate the proton and neutron drift coefficients to the diffusion
coefficients and the associated driving forces, in terms of the Einstein relations as follows
\cite{randrup1979,merchant1982},
\begin{align} \label{eq40}
\nu_{n}=-\frac{D_{NN}}{T}\frac{\partial U}{\partial N_{1}}=D_{NN}&\left[-\alpha\sin\phi\left(\tilde{z}\cos\phi-\tilde{n}\sin\phi\right)\right.\nonumber\\
&\left.+\beta\cos\phi\left(z\sin\phi+n\cos\phi\right)\right],
\end{align}
and
\begin{align} \label{eq41}
\nu_{z}=-\frac{D_{ZZ} }{T}\frac{\partial U}{\partial Z_1}=D_{ZZ}&\left[+\alpha\cos\phi\left(\tilde{z}\cos\phi-\tilde{n}\sin\phi\right)\right.\nonumber\\
&\left.+\beta\sin\phi\left(z\sin\phi+n\cos\phi\right)\right].
\end{align}
\begin{figure}[!hpt]
%\vspace{0.2cm}
\includegraphics*[width=8.5cm]{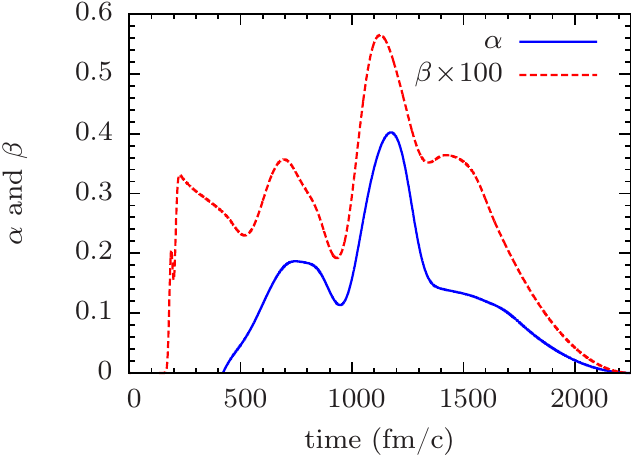}
%\vspace{0.2cm}
\caption{(color online) Curvature parameters $\alpha$ and $\beta$ as a function of time in the
$^{48}\mathrm{Ca}+{}^{238}\mathrm{U}$ collisions in the tip geometry of the uranium with the bombarding energy
$E_{\text{c.m.}} =193$~MeV at the impact parameter $b=2.8$~fm.}
\label{fig5}
\end{figure}
Here the temperature $T$ factor is absorbed into curvature coefficients $\alpha$ and $\beta$,
consequently temperature does not appear as a parameter in the description. We can determine
$\alpha$ and $\beta$ by matching the mean values of neutron and proton drift coefficients obtained
from the TDHF solutions. In this manner, microscopic description of the collision geometry and
details of the dynamical effects are incorporated into the drift coefficients. As an example, Fig.
\ref{fig5} illustrates the curvature parameters $\alpha$ and $\beta$ as a function of time in the
$^{48}\mathrm{Ca}+{}^{238}\mathrm{U}$ collisions in the tip geometry of the uranium with the bombarding energy
$E_{\text{c.m.}} =193$~MeV at the impact parameter $b=2.8$~fm.

\begin{figure}[!hpt]
%\vspace{0.2cm}
\includegraphics*[width=8.5cm]{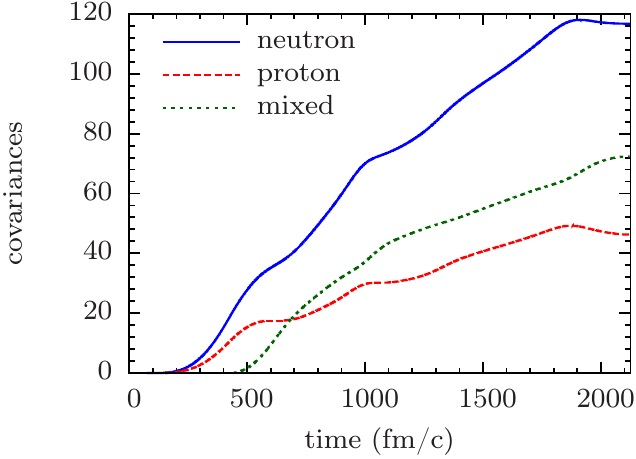}
%\vspace{0.2cm}
\caption{(color online) Co-variances as a function of time in the $^{48}\mathrm{Ca}+{}^{238}\mathrm{U}$ collisions
in the tip geometry of the uranium with the bombarding energy $E_{\text{c.m.}} =193$~MeV at the
impact parameter $b=2.8$~fm.}
\label{fig6}
\end{figure}

The curvature parameters are positive as expected from the potential energy surface, but as a result
of the quantal effects arising mainly from the shell structure, they exhibit fluctuations as a
function of time. Time dependence can also be viewed as a dependence on the relative distance
between ions. In differential equations (\ref{eq17}-\ref{eq19}) for co-variances, we need the
derivatives of drift coefficients with respect to proton and neutron numbers of projectile-like
fragments. As a great advantage of this approach, we can easily calculate these derivatives from
drift coefficients to yield,
\begin{align} \label{eq42}
\frac{\partial\nu_{n}}{\partial N_{1}} =-D_{NN}\left(\alpha\sin^{2}\phi+\beta\cos^{2}\phi\right),
\end{align}
\begin{align} \label{eq43}
\frac{\partial\nu_{z}}{\partial Z_{1}} =-D_{ZZ}\left(\alpha\cos^{2}\phi+\beta\sin^{2}\phi\right),
\end{align}
\begin{align} \label{eq44}
\frac{\partial\nu_{n}}{\partial Z_{1}}=-D_{NN}\left(\beta-\alpha\right)\sin\phi\cos\phi,
\end{align}
\begin{align} \label{eq45}
\frac{\partial\nu_{z}}{\partial N_{1}}=-D_{ZZ}\left(\beta-\alpha\right)\sin\phi\cos\phi.
\end{align}
The curvature parameter $\alpha$ perpendicular to the beta stability valley is much larger than the
curvature parameter $\beta$ along the stability valley. Consequently, $\beta$ does not have an
appreciable effect on the derivatives of the drift coefficients. We determine the co-variances
$\sigma_{NN}^{2}(t)$, $\sigma_{ZZ}^{2}(t)$ and $\sigma_{ZZ}^{2}(t)$ for each impact parameterby
solving the coupled differential equations (\ref{eq17}-\ref{eq19}) with the initial conditions
$\sigma_{NN}^{2}(0)=0$, $\sigma _{ZZ}^{2}(0)=0$ and $\sigma_{ZZ}^{2}(0)=0.$ As an example Fig.
\ref{fig6} shows the co-variances as a function of time in the $^{48}\mathrm{Ca}+{}^{238}\mathrm{U}$ collisions in
the tip geometry of the uranium with the bombarding energy $E_{\text{c.m.}} =193$~MeV at the impact
parameter $b=2.8$~fm. The variance of the fragment mass distribution $\sigma_{AA}^{2}(t)$ is
determined as
\begin{align} \label{eq46}
\sigma_{AA}^{2}(t)=\sigma_{NN}^{2}(t)+\sigma_{ZZ}^{2}(t)+2\sigma_{NZ}^{2}(t).
\end{align}

\section{Fragment mass distribution in $^{48}\mathrm{Ca}+{}^{238}\mathrm{U}$ collisions}
\label{results}
\begin{figure}[!hpt]
%\vspace{0.2cm}
\includegraphics*[width=8.5cm]{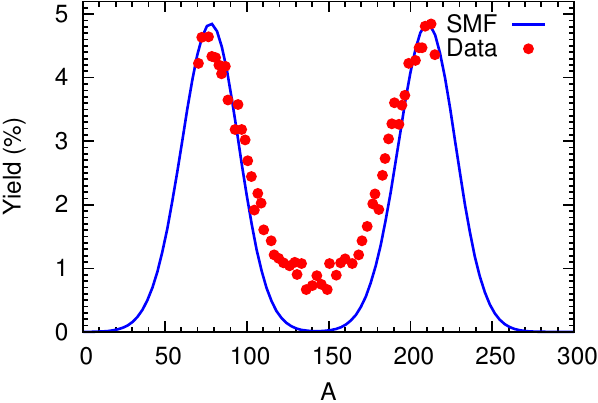}
%\vspace{0.2cm}
\caption{(color online) Primary fragment yield in the $^{48}\mathrm{Ca}+{}^{238}\mathrm{U}$ collisions with the
bombarding energy $E_{\text{c.m.}} =193$~MeV and comparison with data. The solid line is the result
of Eq.~\eqref{eq49}.} \label{fig7}
\end{figure}

\begin{figure}[!hpt]
%\vspace{0.2cm}
\includegraphics*[width=8.5cm]{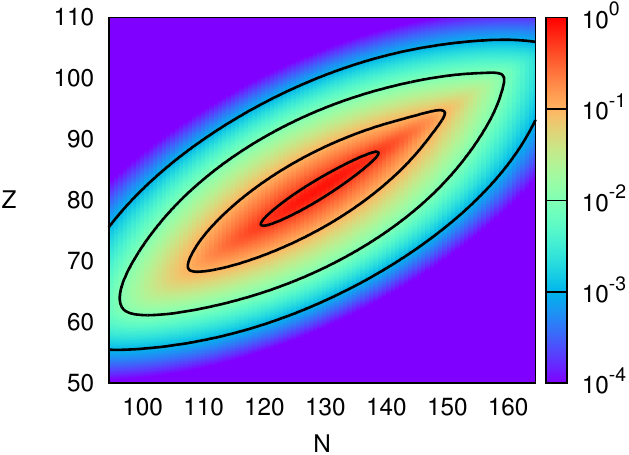}
%\vspace{0.2cm}
\caption{(color online) Calculated primary fragment production cross-section for the $^{48}\mathrm{Ca}+{}^{238}\mathrm{U}$
 collisions with the bombarding energy $E_{\text{c.m.}} =193$~MeV in $(N,Z)$ plane in units
of millibarn.}
\label{fig8}
\end{figure}

In the computation of the production cross-sections of fragments as a function of the charge and
mass of the fragments, we need to include all relevant impact parameters (or the initial orbital
angular momenta) and as well as an average of fragment probabilities over all possible orientations
of the target nucleus $^{238}$U. In the experimental investigations of Kozulin~\textit{et al.}
\cite{kozulin2014b} for the $^{48}\mathrm{Ca}+{}^{238}\mathrm{U}$ system, the detectors are placed between the
angles $+64^\circ$ and $-64^\circ$ with $\pm 10^\circ$ acceptance range in the laboratory frame. We
consider the data collected at the bombarding energy of $E_\text{c.m.}=193$~MeV which corresponds to
$E_{lab}=232$~MeV. In order to determine the dominant geometry of the target nucleus, we consider three
perpendicular configurations of the target which includes ``tip'', ``side-p'' and ``side-s''
orientations. In the ``tip'' orientation the symmetry axis of uranium is parallel to the beam
direction. In the ``side-p'' and ``side-s'' orientations the symmetry axis of uranium is
perpendicular to the beam direction as well as parallel and perpendicular to the reaction plane,
respectively. For a given impact parameter, the laboratory scattering angles $\theta_{1}^{lab}$,
$\theta_{2}^{lab}$ of the fragments are related to the center of mass scattering angle
$\theta_{\text{c.m.}}$ according to,
\begin{align} \label{eq47}
\tan\theta_{1}^{lab}=\frac{\sin\theta_\text{c.m.}}{\sqrt{\frac{A_{1}^{i} A_{1}^{f} }{A_{2}^{i} A_{2}^{f}}\frac{E_{\text{c.m.}}}{TKE}}+\cos\theta_\text{c.m.}},
\end{align}
and
\begin{align} \label{eq48}
\tan \theta _{2}^{lab} =\frac{\sin\theta_\text{c.m.}}{\sqrt{\frac{A_{1}^{i} A_{2}^{f} }{A_{1}^{f} A_{2}^{i}}\frac{E_{\text{c.m.}} }{TKE}}-\cos\theta_\text{c.m.}}     .
\end{align}
Here, $A_{1}^{i}$, $A_{2}^{i}$ and $A_{1}^{f}$, $A_{2}^{f}$ denote the initial and final mass
numbers of the fragments, and $TKE$ is the total kinetic energy of the fragments after the
collision. We calculate the scattering angles the bombarding energy $E_{\text{c.m.}} =193$~MeV with
different impact parameters for different geometries employing the TDHF description. We find that at
this bombarding energy, ``side-p'' and ``side-s'' configurations do not lead to the experimental
scattering angle range with any impact parameter. Therefore, we assume that the dominant
contribution to the experimental range arise from the ``tip'' configuration of the uranium nucleus.
This assumption is supported by the recent investigation of~\cite{itkis2011}. We find that the
collisions in the ``tip'' orientation with the initial orbital angular momenta interval $40\,\hbar
\le \ell\le 62\,\hbar$ reaches to the experimental acceptance range. Tables~\ref{tab1} and
\ref{tab2} present the results of the TDHF calculations in the ``tip'' orientation. Table~\ref{tab1}
shows the initial orbital angular momentum $\ell_{i}$, the corresponding impact parameter $b_{i}$,
the final orbital angular momentum $\ell_{f}$, the final average total kinetic energy $TKE$, the
average total excitation energy $E^{*}$, the center of mass scattering angle $\theta_{\text{c.m.}}$,
and scattering angles $\theta_{1}^{lab}$, $\theta_{2}^{lab}$ of the fragments in the lab frame,
respectively. We calculate the mean excitation energies employing the expression,
$E^{*}=E_{\text{c.m.}} +Q-TKE$, where $Q$ represents the $Q$-value of the channel. Table~\ref{tab2}
shows the mass and charge numbers $A_{1}^{f}$, $A_{2}^{f}$, $Z_{1}^{f}$, $Z_{2}^{f}$ of the final
fragments, co-variances $\sigma_{NN}^{2}$, $\sigma_{ZZ}^{2}$, $\sigma_{NZ}^{2}$ for each orbital
angular momentum.
\begin{table}[h]
\caption{The impact parameter $b_{i}$, the final orbital angular momentum$l_{f}$, the final average
total kinetic energy $TKE$, the average total excitation energy $E^{*}$ and scattering angles
corresponding to the initial orbital angular momentum $\ell_{i}$.}
\label{tab1}
\begin{ruledtabular}
\begin{tabular}{|c|c|c|c|c|c|c|c|}
\hline
$\ell_i\,$($\hbar$) & b$_i\,$(fm) & $\ell_f\,$($\hbar$) & TKE$\,$(MeV) & E$^*$(MeV) & $\theta_{c.m.}$ & $\theta_1^{lab}$ & $\theta_2^{lab}$ \\
\hline
40 & 2.10 & 36.3 & 200.6 & 76.0 & 100.5 & 73.6 & 47.5 \\
42 & 2.20 & 35.8 & 198.4 & 78.2 & 100.2 & 73.2 & 47.6 \\
44 & 2.31 & 38.3 & 193.5 & 82.3 & 96.0 & 69.4 & 49.6 \\
46 & 2.41 & 36.8 & 194.2 & 81.6 & 92.3 & 66.3 & 52.2 \\
48 & 2.51 & 38.1 & 195.3 & 80.5 & 90.5 & 65.0 & 53.2 \\
50 & 2.62 & 40.5 & 193.0 & 78.7 & 87.5 & 62.4 & 54.9 \\
52 & 2.72 & 45.0 & 195.4 & 80.4 & 83.1 & 59.0 & 58.3 \\
54 & 2.82 & 46.8 & 199.2 & 76.6 & 80.8 & 57.3 & 60.2 \\
56 & 2.93 & 46.1 & 198.2 & 77.6 & 76.8 & 54.2 & 62.7 \\
58 & 3.03 & 47.6 & 190.6 & 81.1 & 73.5 & 51.4 & 64.1 \\
60 & 3.14 & 50.5 & 184.1 & 80.9 & 78.7 & 55.1 & 59.1 \\
62 & 3.24 & 49.8 & 180.4 & 76.8 & 88.1 & 62.4 & 52.4 \\
\end{tabular}
\end{ruledtabular}
\end{table}
\begin{table}[h]
\caption{The mass and charge numbers $A_{1}^{f}$, $A_{2}^{f}$, $Z_{1}^{f}$, $Z_{2}^{f}$ of the final
fragments, co-variances $\sigma_{NN}^{2}$, $\sigma_{ZZ}^{2}$, $\sigma_{NZ}^{2}$ for each initial
orbital angular momentum $\ell_{i}$.}
\label{tab2}
\begin{ruledtabular}
\begin{tabular}{|c|c|c|c|c|c|c|c|}
\hline
$\ell_i\,$($\hbar$) & A$_1^{f}$ & Z$_1^{f}$ & A$_2^{f}$ & Z$_2^{f}$ & $\sigma^2_{NN}$ & $\sigma^2_{ZZ}$ & $\sigma^2_{NZ}$ \\
\hline
40 & 78.4 & 31.5 & 207.6 & 80.5 & 159.3 & 67.4 & 75.2 \\
42 & 77.6 & 31.3 & 208.4 & 80.7 & 144.8 & 58.4 & 77.2 \\
44 & 76.7 & 31.0 & 209.3 & 81.0 & 147.4 & 62.3 & 63.0 \\
46 & 77.3 & 31.1 & 208.7 & 80.9 & 139.6 & 57.7 & 75.0 \\
48 & 76.6 & 30.8 & 209.4 & 81.2 & 149.1 & 66.4 & 59.7 \\
50 & 76.2 & 30.6 & 209.8 & 81.4 & 135.9 & 54.3 & 71.7 \\
52 & 77.4 & 31.1 & 208.6 & 80.9 & 154.3 & 65.5 & 64.5 \\
54 & 77.7 & 31.3 & 208.3 & 80.7 & 116.8 & 46.3 & 72.4 \\
56 & 76.8 & 31.0 & 209.2 & 81.0 & 116.0 & 45.7 & 72.1 \\
58 & 76.3 & 30.7 & 209.7 & 81.3 & 112.4 & 44.1 & 70.0 \\
60 & 73.5 & 29.8 & 212.5 & 82.2 & 100.5 & 40.0 & 60.9 \\
62 & 71.7 & 29.0 & 214.3 & 83.0 & 76.8 & 30.1 & 47.2 \\
\end{tabular}
\end{ruledtabular}
\end{table}
We evaluate the mass distributions of the primary fragments as the initial angular momentum weighted
average of the Gaussian functions given in Eq.~\eqref{eq23},
\begin{align} \label{eq49}
P(A)=\frac{\eta}{\sum_\ell(2\ell+1)}\sum_{\ell}(2\ell+1)&\left[P_{\ell}(A-A_{1,\ell})\right.\nonumber\\
 &\left.\;+P_{\ell}(A-A_{2,\ell})\right],
\end{align}
where the first and second Gaussians describe the mass distribution of the projectile-like and the target-like
fragments with the same dispersions $\sigma_{AA}(\ell)$ and the mean values $A_{1,\ell}$, $A_{2,\ell}$, respectively.
In this expression $\eta$ is a normalization constant and the summations run over angular momentum range shown in
Table~\ref{tab1}, $40\le \ell\le 62$,  which approximately corresponds to the data collected in the
experiment reported by Kozulin \textit{et al.}~\cite{kozulin2014b}. We determine the normalization constant
$\eta$ by matching the peak value of the experimental yield at $A=210$ to give a value $\eta=214$
for the normalization. The normalization constant $\eta$ determines the integrated yield between the
peak values $76\le A\le 210$ of the calculated distribution function. In order to make a comparison
with the data, we calculate the area under the data curve within the same interval $76\le A\le 210$
which gives a value of $280$ for the experimental yield. This experimental yield includes the
totally relaxed events within the interval $76\le A\le 210$ and also includes the fusion-fission
events. Since the deep inelastic events are excluded, there are no data points outside the range
$70\le A\le 220$ in Fig.~\ref{fig7}. On the other hand, the quantal diffusion calculations presented
here includes the totally relaxed as well as the deep-inelastic events, but not the fusion-fission
events.
Comparing the integrated yield with the experimental yield in the interval $76\le A\le 210$, the
calculation predicts an integrated yield of $280-214=66$, which is about $23$\% of the total yield,
for the fusion-fission events.  It is also possible to calculate the cross sections for the primary
fragments as a function of neutron and proton numbers by employing Eq.~\eqref{eq22} in discrete
form,
\begin{align} \label{eq50}
\sigma(N,Z)=\frac{\pi\hbar^{2}}{\mu E_{\text{c.m.}}}\sum_{\ell_i}(2\ell_i+1)P_{\ell_i}(N,Z) .
\end{align}
Here, we express the cross-section in terms of the initial orbital angular momenta, rather than the
impact parameters. The summation is over the initial orbital angular momenta listed in Table
\ref{tab1} and $\mu$ is the reduced mass of the projectile and target nuclei. Figure \ref{fig8}
shows the contour plots of the calculated cross-section for producing primary target-like fragments,
in $(N,Z)$ plane in units of millibarn, in the $^{48}\mathrm{Ca}+{}^{238}\mathrm{U}$ collisions with the bombarding
energy $E_{\text{c.m.}} =193$~MeV. Experimental data is not available in~\cite{kozulin2014b} to
compare with the calculated cross-sections.

\section{Conclusions}
\label{conclusions}
We present a quantal diffusion description for multi-nucleon exchange mechanism in dissipative
heavy-ion collisions in the di-nuclear regime.  The diffusion description is deduced by employing
the relevant macroscopic variables in the SMF approach.
In the SMF approach
the collision dynamics is described by an ensemble of mean-field events. The initial conditions for
the events are specified by the quantal and thermal fluctuations in the initial state. In the
di-nuclear regime of the collision, reaction predominantly occurs by nucleon exchange through the
window between the projectile and target nuclei. It is possible to define the neutron and the proton
numbers of projectile-like and target-like fragments, relative momentum and other macroscopic
variables in each event with the help of the window dynamics by integrating the relevant quantities
in both side of the window over the TDHF density. The SMF approach gives rise to Langevin
description for the evolution of macroscopic variables. The Langevin description is equivalent to
the Fokker-Plank transport equation for distribution function of the macroscopic variables. The
transport approach is characterized by diffusion and drift coefficients for macroscopic variable. In
this study, we consider charge and mass asymmetry as the macroscopic variables and drive analytical
quantal expressions for the associated transport coefficients. These transport coefficient are
determined entirely in terms of the mean-field properties provided the solutions of the TDHF
equations. The description of mean values and the fluctuation of the macroscopic variables are
determined by the set of occupied single-particle wave functions of the TDHF approach. This
important result is a reflection of the fluctuation-dissipation relation of the non-equilibrium
quantum statistical mechanics. Quantal diffusion description includes the full geometry of the
collision dynamics and does not involve any adjustable parameter other than the Skyrme parameters of
the TDHF.

As a first application, we applied the quantal diffusion approach to study multi-nucleon
transfer in the $^{48}\mathrm{Ca}+{}^{238}\mathrm{U}$ collisions with the bombarding energy $E_{\text{c.m.}} =193$
MeV. During the long interaction times, of the order of $2000$~fm/c, the di-nuclear system drift
toward symmetry by transferring nearly $20$ neutrons and $10$ protons to the projectile. The large
drift is accompanied with a broad charge and mass distribution with a mass dispersion in the order
of $18$ atomic mass unit. We have calculated the cross-sections for produced fragments as a function of
neutron and proton number, as well as the mass distributions of the primary fragments. In this work, we don't
carry out the de-excitation calculations. However, because of the relatively low excitation energies
of the fragments, we expect de-excitation mechanism to not not alter the primary fragment
distributions appreciably. We analyze the data for the $^{48}\mathrm{Ca}+{}^{238}\mathrm{U}$ collisions at the
bombarding energy $E_{\text{c.m.}} =193$~MeV published by Kozulin et al.~\cite{kozulin2014b}.
The calculations provide a good description of the measured fragment mass distribution.

\begin{acknowledgments}
S.A. gratefully acknowledges the IPN-Orsay and the Middle East Technical University for warm
hospitality extended to him during his visits. S.A. also gratefully acknowledges useful discussions
with D. Lacroix, and very much thankful to F. Ayik for continuous support and encouragement. This
work is supported in part by US DOE Grant Nos. DE-SC0015513 and DE-SC0013847,
and in part by TUBITAK Grant No. 117F109.
\end{acknowledgments}

\appendix
\section{Window dynamics}
We can determine the orientation of symmetry axis of the di-nuclear system with the help of the mass
quadruple moment of the di-nuclear system.  In the center of mass system, the element of the
quadrupole matrix is given by
\begin{align}\label{a1}
 Q_{ij} =3\sigma_{ij}-\left(\sum_{k}\sigma_{kk}\right)\delta_{ij},
\end{align}
here $i,j,k$ indices take values $1,2,3$ and $x_{1},x_{2},x_{3}\to x,y,z$. In this expression the
elements of the sigma matrix are defined in terms of the position co-variances as,
\begin{align}\label{a2}
 \sigma_{ij}(t)=&\sum_{h}\langle\phi_{h} (t)|x_{i} x_{j}|\phi_{h}(t)\rangle\nonumber\\
 &-\sum_{h,h'}\langle\phi _{h}(t)|x_{i} | \phi _{h} (t)\rangle\langle\phi _{h'} (t)|x_{j} |\phi _{h'} (t)\rangle.
\end{align}
We can determine the direction of the symmetry axis, with the elements of the quadrupole matrix in
reaction plane which is defined by $z=0$. In this case, we just need to diagonalize the $2\times2$
reduced quadrupole matrix on the reaction plane with elements $Q_{xx}$, $Q_{xy}$, $Q_{yx}$,
$Q_{yy}$. The eigenvectors $\vec{E}_{+}$ and $\vec{E}_{-}$ of the quadrupole matrix specify the
principal axes of the di-nuclear mass distribution on the reaction plane. These eigenvectors have
the following form,
\begin{align}
 \vec{E}_{\pm}=\left(\begin{array}{c} {Q_{\pm } } \\ {1} \end{array}\right),
\end{align}
with components  given by
\begin{align}
 Q_{\pm } =\frac{Q_{xx} -Q_{yy} \pm \sqrt{(Q_{xx} -Q_{yy} )^{2} +4Q_{xy}^{2} } }{2Q_{xy} }.
\end{align}
The eigenvalues $\Gamma_{\pm}$ corresponding to $Q_{\pm}$ are given by,
\begin{align}
 \Gamma _{\pm } =\frac{Q_{xx} +Q_{yy} \pm \sqrt{(Q_{xx} -Q_{yy} )^{2} +4Q_{xy}^{2} } }{2}.
\end{align}
The beam direction as taken in the $x$ direction in the fixed coordinate system $(x,y)$. The
eigenvectors of the quadrupole matrix define a rotated orthogonal system in the reaction plane with
axes pointing along $\vec{E}_{+}$ and $\vec{E}_{-}$ directions.  We note that the eigenvalue $\Gamma
_{+}$ is larger than the eigenvalue $\Gamma _{-}$. The eigenvector $\vec{E}_{+}$ associated with the
large eigenvalue $\Gamma _{+}$ specifies the direction of the symmetry axis of the di-nuclear
system. The angle $\theta$ between the positive direction of the x-axis and the direction of
$\vec{E}_{+} =(Q_{+} ,1)$ is determined by
\begin{align}
 \tan \theta =\frac{1}{Q_{+}}=\frac{2Q_{xy}}{Q_{xx} -Q_{yy} +\sqrt{(Q_{xx} -Q_{yy} )^{2} +4Q_{xy}^{2} } }
\end{align}
Using the trigonometric identity,
\begin{align}
 \tan \theta =\frac{\tan 2\theta }{1+\sqrt{1+\tan ^{2} 2\theta } } \text{,}\quad-\frac{\pi }{2} <\theta <+\frac{\pi }{2}
\end{align}
and
\begin{align}
 \tan \theta =\frac{\tan 2\theta }{1-\sqrt{1+\tan ^{2} 2\theta } } \text{,}\quad+\frac{\pi }{2} <\theta <+\frac{3\pi }{2}
\end{align}
we can express the angle $\theta$ symmetry axis in terms of the elements of the elements of the
quadrupole or the elements of sigma matrix as,
\begin{align}
 \tan 2\theta =\frac{2Q_{xy} }{Q_{xx} -Q_{yy} } =\frac{2\sigma _{xy} }{\sigma _{xx} -\sigma _{yy} }
\end{align}
which applies to the entire range specified in (A7) and (A8).

\section{Analysis of the closure relation}
We re-write Eq.~\eqref{eq32} as,
\begin{align}
 \sum _{a\in P,h\in T}A_{ah}^{\alpha } (t)A_{ah}^{*\alpha } (t') =\sum _{h\in T}\int & d^{3} R d^{3} r\delta (\vec{r}+\vec{u}_{h} \tau )\nonumber\\
 &\times W_{h}^{\alpha } (\vec{r}_{1} ,t)W_{h}^{*\alpha } (\vec{r}_{2} ,t'),
\end{align}
in which we introduce the coordinate transformation,
\begin{align}
 \vec{R}=(\vec{r}_{1}+\vec{r}_{2})/2 \text{,}\quad\vec{r}=\vec{r}_{1} -\vec{r}_{2},
\end{align}
and its reverse as
\begin{align}
 \vec{r}_{1} =\vec{R}+\vec{r}/2 \text{,}\quad\vec{r}_{2} =\vec{R}-\vec{r}/2.
\end{align}
For clearness, we present quantities $W_{h}^{\alpha } (\vec{r}_{1} ,t)$ and $W_{h}^{*\alpha } (\vec{r}_{2} ,t')$  here again,
\begin{align}
 W_{h}^{\alpha } (\vec{r}_{1} ,t)&=\frac{\hbar }{m} g(x'_{1} )\nonumber\\
 &\;\;\;\times\left(\hat{e}\cdot \vec{\nabla }_{1} \Phi _{h}^{\alpha } (\vec{r}_{1} ,t)-\frac{x'_{1} }{2\kappa ^{2} } \Phi _{h}^{\alpha } (\vec{r}_{1} ,t)\right),
\end{align}
and
\begin{align}
 W_{h}^{*\alpha } (\vec{r}_{2} ,t')&=\frac{\hbar }{m} g(x'_{2} )\nonumber\\
 &\;\;\;\times\left(\hat{e}\cdot \vec{\nabla }_{2} \Phi _{h}^{*\alpha } (\vec{r}_{2} ,t')-\frac{x'_{2} }{2\kappa ^{2} } \Phi _{h}^{*\alpha } (\vec{r}_{2} ,t')\right).
\end{align}
Because of the delta function in the integrant of Eq.~(B1), we make the substitution  $\vec{r}=-\vec{u}_{h}^{\alpha } (\vec{R},T)\tau$ in the wave functions and introduce the backward diabatic shift to obtain,
\begin{align}
 \Phi _{h}^{\alpha } (\vec{R}+\vec{r}/2,t)=\Phi _{h}^{\alpha } (\vec{R}-\vec{u}_{h}^{\alpha } \tau /2,t)\approx \Phi _{h}^{\alpha } (\vec{R},T),
\end{align}
and
\begin{align}
 \Phi _{h}^{\alpha } (\vec{R}-\vec{r}/2,t')=\Phi _{h}^{\alpha } (\vec{R}+\vec{u}_{h}^{\alpha } \tau /2,t')\approx \Phi _{h}^{\alpha } (\vec{R},T).
\end{align}
The local flow velocity of the wave function $\Phi _{h}^{\alpha } (\vec{R},T)$ is calculated in the standard manner,
\begin{align}
 \vec{u}_{h}^{\alpha } (\vec{R},T)=\frac{\hbar }{m} \frac{1}{|\Phi _{h}^{\alpha } (\vec{R},T)|^{2} }\text{Im}\left(\Phi _{h}^{*\alpha } (\vec{R},T)\nabla \Phi _{h}^{\alpha } (\vec{R},T)\right),
\end{align}
with $T=(t+t')/2=t-\tau /2$. We write the product of Gaussian factors as
\begin{align}
 g(x'_{1} )g(x'_{2} )=\tilde{g}(X')\tilde{G}(x'),
\end{align}
with
\begin{align}
 \tilde{g}(X')=\frac{1}{\sqrt{\pi }\kappa}\exp \left[-\left(\frac{X'}{\kappa } \right)^{2} \right],
\end{align}
and
\begin{align}
 \tilde{G}(x')=\frac{1}{\sqrt{4\pi }\kappa }\exp \left[-\left(\frac{x'}{2\kappa } \right)^{2} \right].
\end{align}
In these Gaussians following the transformations below Eq.~\eqref{eq4}, the coordinates in the
rotating frame are expressed in terms of the coordinates in the fixed frame as, $X'=(X-x_{0} )\cos
\theta +(Y-y_{0} )\sin \theta$ and $x'=x\cos \theta +y\sin \theta$. Carrying out the product of the
factors and making the substitution $\vec{r}=-\vec{u}_{h}^{\alpha } (\vec{R},T)\tau$, Eq.~(B1)
becomes,
\begin{align}
&\sum_{a\in P,h\in T}A_{ah}^{\alpha } (t)A_{ah}^{*\alpha } (t')=\left(\frac{\hbar }{m} \right)^{2} \sum _{h\in T}\int d^{3} R  \tilde{g}(X')\frac{G_{T}^{h} (\tau )}{|u_{\bot }^{h} (\vec{R},T)|} \nonumber\\
&\qquad\times\left[|\hat{e}\cdot \vec{\nabla }\Phi _{h}^{\alpha } (\vec{R},T)|^{2} +\frac{X'^{2} -(u_{\bot }^{h} \tau /2)^{2} }{4\kappa ^{4} } |\Phi _{h}^{\alpha } (\vec{R},T)|^{2} \right.\nonumber\\
&\left.\qquad\qquad-\frac{X'}{2\kappa ^{2} } \hat{e}\cdot \vec{\nabla }(|\Phi _{h}^{\alpha } (\vec{R},T)|^{2} )\right].
\end{align}
Here, $u_{\bot }^{h} (\vec{R},T)$ represents the component of the nucleon (proton or neutron) flow
velocity perpendicular to the window $u_{\bot }^{h} (\vec{R},T)=\hat{e}\cdot \vec{u}_{T}^{h}
(\vec{R},T)$ and $G_{h} (\tau )$ indicates the memory kernel,
\begin{align}
 G_{T}^{h} (\tau )=\frac{1}{\sqrt{4\pi } } \frac{1}{\tau _{T}^{h} } \exp [-(\tau /2\tau _{T}^{h} )^{2} ],
\end{align}
with the memory time $\tau _{T}^{h} =\kappa /|\vec{u}_{\bot }^{h} |$. In this expression
$\tilde{g}(X')$ is sharp as Gaussian smoothing function centered on the window with a dispersion
$\kappa =0.5$~fm due to the fact that $\tilde{g}(X')$ is centered at $X'=0$, the third term in Eq.
(B12) is nearly zero. In the second term, after carrying out an average over the memory, the factor
in the middle becomes,
\begin{align}
 X'^{2} -(u_{x}^{h} \tau /2)^{2} \to X'^{2} -(\kappa /2)^{2}.
\end{align}
Since Gaussian $\tilde{g}(X')$ is sharply peaked around $X'=0$ with a variance $(\kappa /2)^{2}$,
the second terms in Eq.~(B12) is expected to be very small, as well. Neglecting the second and third
terms, Eq.~(B1) becomes,
\begin{align}
 \sum _{a\in P,h\in T}A_{ah}^{\alpha } (t)A_{ah}^{*\alpha } (t') =\left(\frac{\hbar }{m} \right)^{2} & \sum _{h\in T}\int d^{3} R\,  \tilde{g}(X')\frac{G_{T}^{h} (\tau )}{|u_{\bot }^{h} (\vec{R},T)|}\nonumber\\
 &\quad\times|\hat{e}\cdot \vec{\nabla }\Phi _{h}^{\alpha } (\vec{R},T)|^{2}.
\end{align}

In the continuation, we express the wave functions in terms of its magnitude and its phase as
\cite{gottfried1966},
\begin{align}
 \Phi _{h}^{\alpha}(\vec{R},T)=|\Phi_{h}^{\alpha}(\vec{R},T)|\exp\left(iQ_{h}^{\alpha}(\vec{R},T)\right).
\end{align}
The phase factor $Q_{h}^{\alpha } (\vec{R},T)$ behaves as the velocity potential of the flow
velocity of the wave. Using the definition given by Eq.~(B8), we observe that the flow velocity is
given by $\vec{u}_{h}^{\alpha } (\vec{R},T)=(\hbar /m)\vec{\nabla }Q_{h}^{\alpha } (\vec{R},T)$. In
the vicinity of the window, in the perpendicular direction, the phase factors varies faster than the
magnitude of the wave functions. Neglecting the variation of the magnitude $|\Phi _{h} (\vec{R},T)|$
in the vicinity of the window, we can express the gradient of the wave function in Eq.~(B12) as,
\begin{align}
 \hat{e}\cdot \vec{\nabla }\Phi _{h}^{\alpha } (\vec{R},T)\approx i\Phi _{h}^{\alpha } (\vec{R},T)\left(\hat{e}\cdot \vec{\nabla }Q_{h}^{\alpha } (\vec{R},T)\right),
\end{align}
where the quantity inside the parenthesis is the component of the nucleon flow velocity
perpendicular to the window. As a result, Eq.~(B1) becomes,
\begin{align}
 \sum _{a\in P,h\in T}A_{ah}^{\alpha } (t)A_{ah}^{*\alpha } (t') =\int d^{3} R \tilde{g}(X')\tilde{J}_{\bot ,\alpha }^{T} (\vec{R},t-\tau /2).
\end{align}
Here, $\tilde{J}_{\bot ,\alpha }^{T} (\vec{R},t-\tau /2)$ represents the magnitude of the current
densities perpendicular to the window due to each wave functions originating from target and each
term multiplied by the memory kernel,
\begin{align}
 \tilde{J}_{\bot ,\alpha }^{T} (\vec{R},T)=\frac{\hbar }{m} & \sum _{h\in T}G_{T}^{h} (\tau )\nonumber\\
 &\times\left|\text{Im}\left[\Phi _{h}^{*} (\vec{R},T)\left(\hat{e}\cdot \vec{\nabla }\Phi _{h} (\vec{R},T)\right)\right]\right|.
\end{align}

In Eq.~\eqref{eq33} we introduce a further approximation by replacing the individual memory kernels
$G_{T}^{h} (\tau )$ by its average value taken over the hole states,
\begin{align}
 G_{T} (\tau )=\frac{1}{\sqrt{4\pi } \tau_{T}} \exp [-(\tau /2\tau _{T} )^{2} ],
\end{align}
with the memory time determined by the average speed $u_{T}$ by $\tau _{T} =\kappa /|u_{T} (t)|$.
The average value $u_{T}^{h} (t)$ of the flow speed for each hole state across the window is
calculated as $u_{T}^{h} (t)=\int d^{3} R \tilde{g}(X')\hat{e}\cdot \vec{j}_{T}^{h} (\vec{R},t)/\int
d^{3} R \tilde{g}(X')\rho _{T}^{h} (\vec{R},t)$ where $\vec{j}_{T}^{h} (\vec{R},t)$ and $\rho
_{T}^{h} (\vec{R},t)$ denote the current density and density of hole state originating from target,
respectively. The average $|u_{T} (t)|$ is then calculated by taking the mean value of all the flow
speeds $|u_{T}^{h} (t)|$ of the hole states. It is possible to calculate the average speed from
$u_{T} (t)=\int d^{3} R \tilde{g}(X')\hat{e}\cdot \vec{j}_{T} (\vec{R},t)/\int d^{3} R
\tilde{g}(X')\rho _{T} (\vec{R},t)$, where $\rho _{T} (\vec{R},t)$ and $\vec{j}_{T} (\vec{R},t)$ are
the total density and the total current density of the states originating from the target. We expect
both average speeds have nearly the same magnitude.

\bibliography{VU_bibtex_master}

\end{document}